\documentclass[epj]{svjour}
\usepackage[T1]{fontenc}
\usepackage{times}

\usepackage{graphics}

\begin{document}

\title{Atom-optical elements on micro chips}

\author{S. Kraft \and A. G\"{u}nther \and P. Wicke \and B. Kasch \and C. Zimmermann \and J. Fort\'{a}gh}%
\institute{Physikalisches Institut der Universit\"{a}t T\"{u}bingen \\ Auf
der Morgenstelle 14, 72076 T\"{u}bingen, Germany}
\date{Received: date / Revised version: date}

\abstract{We describe the realization of atom-optical elements as
magnetic waveguide potentials, beam splitters and gravitational
traps on a microchip. The microchip was produced by electroplating
gold conductors on an aluminium-oxide substrate. The conductors are
30-150 $\mu$m wide and allow for the generation of waveguides at
large distances to the chip surface, where surface effects are
negligible. We show that these elements can be integrated on a
single chip to achieve complex atom-optical circuits.
\PACS{
      {03.75.Be}{Atom and neutron optics} \and
      {39.25.+k}{Atom manipulation}   \and
      {39.90.+d}{Other instrumentation and techniques for atomic and molecular physics}
    } %
}%
\maketitle
\section{Introduction}

The manipulation of Bose-Einstein condensates by means of
microscopic electromagnets \cite{Ott2001,Hansel2001} has developed
into an intensive  field  of research during the past years. The
vision of matter-wave interferometers and the control of single
atoms \cite{Dorner2003} seems within reach by using micron scaled
potential structures at a chip surface. Current micro fabrication
technologies allow the production of high quality electromagnets
which meet the requirements of controlling the motion of atoms on
the quantum level. On the other hand, some limitations of atom
chips have recently been explored including spin-flip losses near
metallic surfaces \cite{Henkel1999a}, dispersive atom-surface
interactions \cite{Lin2004}, and technical limitations such as
geometrical imperfections of micro electromagnets
\cite{Kraft2002,Esteve2004}. As important as micron scaled
potential barriers  are smooth waveguide potentials in
atom-optical circuits. These enable guiding and splitting of
matterwaves, similar to the manipulation of photons in optical
fibers and beam splitters. It is preferable to form such
waveguides relatively far from the chip surface where the cloud is
less affected by surface effects.

\begin{figure}[h]
  \centering
  \resizebox{\columnwidth}{!}{
  \includegraphics{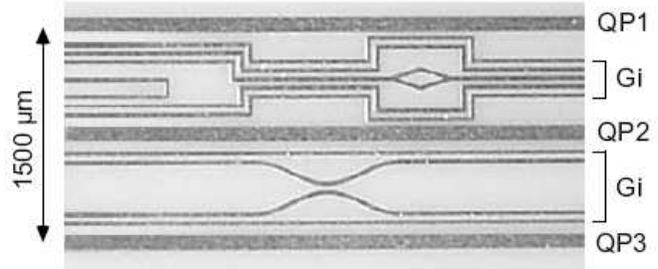}}
  \caption{Complex atom-optical circuit. Waveguide and interferometer geometries as well as
  gravitational traps are arranged between QP1-QP2 and QP2-QP3. The QP conductors
  define a central waveguide from which neighboring microtraps can be loaded adiabatically.}
  \label{setup}
\end{figure}

In this article, we describe magnetic waveguides, beam splitters,
and gravitational traps. These atom-optical elements are all
implemented on a single chip (Fig.~\ref{setup}) currently used in
our groups, which is routinely loaded with Bose-Einstein
condensates \cite{GuentherPRA2005}.
It has been produced by electroplating a 6~$\mu$m thick gold layer
on a 250~$\mu$m thick aluminium-oxide substrate. The minimal width
and spacing  of the conductors is 30 $\mu$m. The chip consists of
three conductors with a constant width of 100 $\mu$m (QP1 -- QP3)
and nine additional conductors with varying width (G\emph{i}).
Perpendicular conductors on the back side of the chip allow
positioning of the trap parallel to the conductors QP
\cite{GuentherPRA2005}. Experimentally, we found that in a pulsed
duty cycle (3s operation time in 60s cycle time) the $100\,\mu$m
and 30 $\mu$m wide conductors can carry a current up to 2A and 1A,
respectively. An atomic cloud, initially confined in a waveguide
potential above QP2  can be adiabatically loaded into the
mictrotraps between QP1 and QP2, as well as between QP2 and QP3.
Neighboring traps can exchange atoms directly, more distant traps
are connected via waveguides.

\section{Highly elongated traps}

A simple waveguide is formed by the circular field of a thin
current carrying conductor superimposed by a homogeneous bias
field \cite{Fortagh1998,Reichel2002,Folman2002}. The two fields
compensate each other along a line parallel to the conductor.
Centered around this line, the magnetic field is well approximated
by a two dimensional quadrupole, and paramagnetic atoms in a low
field seeking state become trapped in radial direction. The
quadrupole channel is characterized by the gradient of the
magnetic field in the radial direction $a_\mathrm{r}$. If an
additional magnetic offset field $B_\mathrm{off}$ is applied
parallel to the waveguide \cite{Pritchard1983}, the radial
confinement becomes harmonic, characterized by the radial
oscillation frequency $\omega_\mathrm{r} = a_\mathrm{r} \cdot
\sqrt{g_\mathrm{F}\mu_\mathrm{B}\mathrm{m_{F}}/(m B_\mathrm{off})}
$, with the Land\'e factor $g_\mathrm{F}$, the mass of the atom
$m$, and the Bohr magneton $\mu_\mathrm{B}$. In this article, the
trap frequencies are calculated for $^{87}$Rb atoms in the F=2,
$\mathrm{m_F}=2$ hyperfine ground state.

\subsection{Waveguides with parallel wires}

\begin{figure}
  \centering
  \resizebox{\columnwidth}{!}{%
  \includegraphics{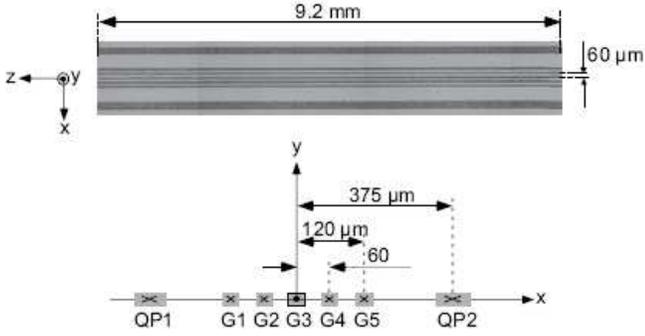}}
  \caption{Waveguide with parallel conductors on a chip. (Top) Microscope image of the chip.
  (Bottom) Conductor geometry. If the center conductor G3 is driven with a current opposite
  in direction to the outer conductors (QP1, G1, G2, G4, G5 and QP2), a waveguide potential
  forms. With pair wise equal currents in QP1-QP2, G1-G5, and G2-G4, the waveguide forms above G3.}
  \label{1dtrap}
\end{figure}

A waveguide potential with parallel on-chip conductors has been
demonstrated \cite{Dekker2000}. Here, we describe the realization
of a waveguide which allows increasing the radial compression by
using additional wires for the bias field. The section of the chip
used to build the waveguide consists of 7 parallel conductors
(Fig.~\ref{1dtrap}). Five conductors of the width 30 $\mu$m are in
the center area (G1 - G5) with centers separated by 60 $\mu$m.
Another two, wider conductors QP1 and QP2 (100 $\mu$m width)
complete the setup. The centers of these outer conductors are  375
$\mu$m from the conductor G3 in the middle.
The current of G3 is opposite in direction to the other
conductors. The waveguide is centered above G3 by pair wise using
the same currents: $I_\mathrm{G2}=I_\mathrm{G4}$,
$I_\mathrm{G1}=I_\mathrm{G5}$ and $I_\mathrm{QP1 } =
I_\mathrm{QP2}$. The circular magnetic field of the center
conductor is then superposed by the bias field arising from the
outer conductors. In this geometry the radial gradient is not only
due to the gradient of the central conductor but also due to a
gradient of the inhomogeneous bias field.

For simplification, let us first assume a waveguide achieved by
the three conductors G2, G3, and G4 of Fig. \ref{1dtrap}. The
separation between the middle of the conductors is $d$. The
magnetic field on the y axis of this configuration is given by the
sum of the fields of the central conductor and of the outer
conductors
\[
B(x=0,y) = - \frac{\mu_0}{2\pi}\frac{I_\mathrm{C}}{y} +
\frac{\mu_0}{\pi} \frac{I_\mathrm{O}y}{d^2 + y^2} \; ,
\]
with $I_\mathrm{C}$ the current in the central conductor G3 and
$I_\mathrm{O}$ the current in each of the outer conductors G2 and
G4. The waveguide formes at a distance $y_0$ to the surface:
\[
y_0 = d \sqrt{\frac{I_\mathrm{C}}{2I_\mathrm{O} - I_\mathrm{C}}}\; .
\]
The radial gradient $a_\mathrm{r}$ is given by the derivative of the
magnetic field at this point. In terms of the current in the outer
conductors $I_\mathrm{O}$ and the distance of the trap center to the
chip $y_0$ the gradient is given by
\[
a_\mathrm{r}(I_\mathrm{O},y_0) =
\frac{\mu_0}{\pi}\frac{I_\mathrm{O}}{d^2} \cdot \frac{1}{(1 +
(y_\mathrm{0}/d)^2)^2} \; .
\]
There are two ways to increase the gradient. One is to increase
the current in the outer conductors $I_\mathrm{O}$. As this
reduces $y_0$, it increases the gradient stronger than linearly.
However, the flow of the dissipated heat sets an upper limit
$I_\mathrm{O,max}$ for this current. The other way is to lower
$y_0$ by decreasing the current in the central conductor
$I_\mathrm{C}$. At $I_\mathrm{C} = 0$, the maximal possible
gradient of
\[
a_\mathrm{r,max} = 4
\frac{\mu_0}{2\pi}\frac{I_\mathrm{O,max}}{d^2}
\]
is reached, in the limit of touching the conductor. This
corresponds to a waveguide build only by two wires with currents
in the same direction \cite{Thywissen1999}.

In the setup shown in Figure~\ref{1dtrap}, additional pairs of
wires contribute to the bias field. The increased bias field
allows higher current in the central conductor and hence increases
the maximal possible gradient at the position of the trap. For the
seven-wire configuration and with currents of 1A in G1, G2, G4, G5
and 2A in QP1, QP2, we calculate a radial gradient of
$a_\mathrm{r}(y_0=0) = 289\,\mathrm{Tm}^{-1}$. With an axial
offset field of 1 G, the radial oscillation frequency is
$\omega_\mathrm{r}(y_0=0) = 2\pi \cdot 36.8$ kHz. Table
\ref{traptable} shows typical values for different currents in the
conductors.
\begin{table}
\begin{center}
\begin{tabular}{|c||c|c|c|c|c|c|}
\hline
$y_0$&$I_\mathrm{QP}$ & $I_\mathrm{G1,G5}$ & $I_\mathrm{G2,G4}$ & $I_\mathrm{G3}$ & $a_\mathrm{r}$ & $\nu_\mathrm{r}$\\
$[\mu\mathrm{m}] $ & $\mathrm{[ A ] }$ & $\mathrm{[ A ] }$ & $\mathrm{[ A ] }$ & $\mathrm{[A]} $ & $\mathrm{[T/m]} $ & $\mathrm{[kHz]}$\\\hline\hline%
100    & 2  & 0.9        &0      & 1        & 27    &  3.5\\\hline%
90    & 2  & 1        &0.044      & 1        & 34    &  4.3\\\hline%
80     & 2     &  1       & 0.164        & 1       & 42  & 5.3\\\hline%
70     & 2     &  1       & 0.31        & 1       & 54  & 6.9\\\hline%
60     & 2     &  1       & 0.5        & 1       & 74  & 9.4\\\hline%
50     & 2     & 1        & 0.773        &  1        & 111  & 14.1\\\hline%
40     & 2     &  1       & 1       & 0.861       & 163  & 20.7\\\hline%
30     & 2     & 1         & 1       & 0.544      & 203 & 25.8\\\hline%
20     & 2     &  1       & 1        & 0.266      & 244  & 31\\\hline%
10     & 2     & 1         & 1       &  0.071   & 277 & 35.2\\\hline
\end{tabular}
\end{center}
\caption{Gradients and radial oscillation frequencies of the
waveguide (Fig.~\ref{1dtrap}) for different distances $y_0$ above
the central conductor G3. To calculate the trap frequencies, a
homogeneous offset field of 1G along the waveguide was assumed.}
\label{traptable}
\end{table}
Note, that the currents in the conductors producing the bias field
are not driven equally. For traps near to the surface, a small
distance of the conductors generating the bias field to the
central conductor is favorable, while for traps far from the
surface, a bigger distance is preferable. More conductors increase
the flexibility for choosing the position of the bias field
generating elements.

\subsection{Folded waveguide configuration}

A chip with a set of parallel conductors as in Fig.~\ref{1dtrap}
allows tight radial confinement even with moderate currents.
However, the waveguide is located at a rather small distance to the
chip surface. Imperfections of the conductor geometry may thus make
the potential of the waveguide irregular leading to undesired
fragmentation of atomic clouds \cite{Kraft2002,Esteve2004}.

\begin{figure}
  \centering
  \resizebox{\columnwidth}{!}{
  \includegraphics{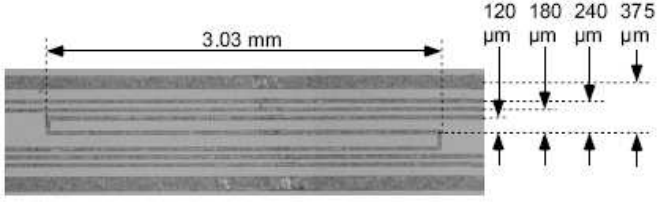}}
  \caption{In the folded alinement, the circular magnetic field of the central
  conductor generates its own bias field. This adds to the
  compression due to the outer conductors and increases the confinement.}
  \label{1dZtrap}
\end{figure}

To achieve a smooth waveguide potential a large surface --
waveguide distance is preferable. When the total number of
conductors is fixed a folded wire configuration
(Fig.~\ref{1dZtrap}) can increase the radial gradient at certain
distances.
In this setup, the folded upper and lower parts of the central
conductor contribute to the bias field of the outer conductors.
The contribution becomes relevant for currents in G3 which are
comparable to the current in the outer conductors and is
negligible for smaller currents. For small currents, i.e. traps
near the surface of the chip, the maximum achievable compression
and gradient are reduced compared to the previous setup
(Fig.~\ref{1dtrap}) because the conductors generating the bias
field are further away from the center. Table \ref{table2}
\begin{table}
\begin{center}
\begin{tabular}{|c||c|c|c|c|c|c|}
\hline
$y_0$&$I_\mathrm{QP}$ & $I_\mathrm{G1,G5}$ & $I_\mathrm{G2,G4}$ & $I_\mathrm{G3}$ & $a_\mathrm{r}$ & $\nu_\mathrm{r}$\\
$[\mu\mathrm{m}] $ & $\mathrm{[ A ] }$ & $\mathrm{[ A ] }$ & $\mathrm{[ A ] }$ & $\mathrm{[A]} $ & $\mathrm{[T/m]} $ & $\mathrm{[kHz]}$\\\hline\hline%
100    & 1.35  & 0        &0      & 1        & 26    &  3.3\\\hline%
90     & 2     & 0.25     &0      & 1       & 36  & 4.5 \\\hline%
80     & 2     &  1       & 0.03        & 1       & 49  & 6.2\\\hline%
70     & 2        & 1        & 0.764   &  1   & 68 & 8.6 \\\hline%
60     & 2     & 1        & 1        &  0.696        & 68  & 8.6\\\hline%
50     & 2     & 1        & 1        &  0.421        & 62  & 7.9\\\hline%
40     & 2     & 1        & 1        &  0.242        & 58  & 7.3\\\hline%
30     & 2     & 1         & 1       & 0.125      & 54 & 6.9\\\hline%
20     & 2     & 1        & 1        &  0.0525        & 52  & 6.6\\\hline%
10     & 2     & 1         & 1       &  0.0128   & 48 & 6.1\\\hline %
\end{tabular}
\end{center}
\caption{Gradients and radial oscillation frequencies of the
waveguide produced by a folded wire configuration
(Fig.~\ref{1dZtrap}) for different distances $y_0$ above the
central conductor G3. To calculate the trap frequencies, a
homogeneous offset field of 1G along the waveguide was assumed.}
\label{table2}
\end{table}
shows gradients and trap frequencies for the same set of distances
as in table 1. The variation of the gradient with the distance is
reduced compared to the setup with straight conductors. At working
distances of 70 - 100 $\mu$m, however, the gradient is larger in
this configuration. At larger distances the solution with straight
conductors produces larger trap frequencies. This is because in
the folded geometry the inner conductors contribute to the bias
field. For large distances it would be preferable to place the
conductors generating the bias field further from the middle
conductor.

Comparing the two geometries shows that folding the central
conductor increases the radial confinement at certain distances.
As part of the bias field is always produced by the inner
conductor one looses flexibility in choosing the field generating
elements as  freely as in the setup with straight conductors. This
leads to a smaller bias field outside of the ideal distance range.

\section{Beam splitters}

A more complex atom optical element than a waveguide is a beam
splitter \cite{Folman2002}, in which the atomic matter wave is
divided into two parts. In atom-optical circuits, the divided
matterwave would propagate in separate waveguides before an
inverse beam splitter would recombine them producing interference.

\begin{figure}[t]
  \centering
  \resizebox{\columnwidth}{!}{
  \includegraphics{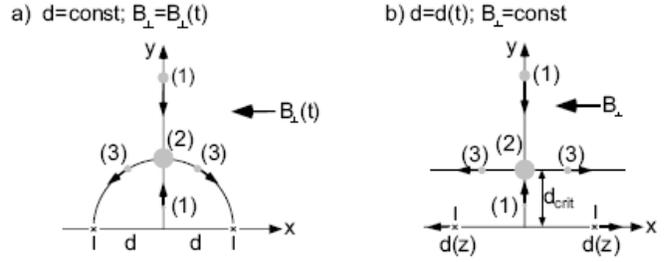}}
  \caption{Beam splitter geometries.  a) Beam splitting by changing the bias field
above two conductors separated by a constant distance. The
trajectories of quadrupole waveguides is shown while the bias field
changes. b) Beam splitting with constant bias field and changing the
distance between the conductors. The trajectories of the waveguide
potentials are shown as explained in the text.}
  \label{beamsplitter1}
\end{figure}

Fig.~\ref{beamsplitter1} shows two different principles to
realize a beam splitter. Both rely on two wires driven with currents
in the same direction and a homogeneous bias field $B_\perp$. The
chip surface is assumed to be horizontal with the gravity parallel
to the surface vector.

\subsection{Beam splitters based on a varying bias field}

The field of two parallel conductors combined with a variable bias
field can be used for splitting an atomic wave function
\cite{Hinds2001}. Fig.~\ref{beamsplitter1}~a) illustrates the
geometry and the trajectories of waveguide potentials for a
varying bias field. Initially, in a small bias field (1), two
quadrupole waveguides are located on the y-axis. While the bias
field is increased, the two waveguides approach each other. At a
critical bias field $B_\mathrm{crit} =
\frac{\mu_0}{2\pi}\frac{I}{d}$ they merge to a hexapole (2) at a
distance $d$ to the surface of the chip. The distance $d$ is half
of the separation between the two conductors. Further increase of
the bias field splits the waveguide into another two quadrupole
waveguides. They move along a half circle to the position of the
conductors (3).

The realization of such a beam splitter in our setup is shown in
Fig.~\ref{1dtrap}. The beam splitter geometry is achieved by
applying currents in G2 and G4 as well as in QP1 and QP2 for
generating the bias field. When the current in these conductor is
increased the bias field increases.

\subsection{Beam splitters based on  varying the distance between the conductors}

Another possible implementation of a beam splitter is shown in
Fig.~\ref{beamsplitter1} b). The bias field stays constant while
the distance $d(z)$ between the conductores varies
\cite{Cassettari2000}.

If the distance $d$ of the two conductors to the y-axis is smaller
than a critical distance $d_\mathrm{crit} =
\frac{\mu_0}{2\pi}\frac{I}{B_\perp}$ (1), the two waveguides lie on
the y-axis ($I$ is the current in both of the conductors). As the
distance between the conductors increases, the waveguides merge
vertically into a single hexapole waveguide at the distance
$d_\mathrm{crit} $ from the surface (2). Further increasing the
distance of the conductors splits the waveguides into two
quadrupoles which separate at constant height above the chip
surface.

\begin{figure}
  \centering
  \resizebox{\columnwidth}{!}{
  \includegraphics{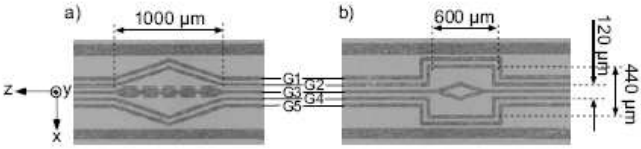}}
  \caption{Realization of beam splitters (and inverse beam splitters) on a
  chip.  a) Interferometers are defined
  by the pairs of conductors G1+G5 and G2+G4. The bias field can be
  generated by QP1 and QP2. In b) the conductors G1,G2, G4 and G5 can be used for
  beam splitters. The separation of
  the waveguides is faster then in a) due to the instantaneous increase of
  the distance of the conductors. The conductor (G3)
  and a bias field generated by the surrounding conductors define another interferometer.
  }
  \label{beamsplitter2}
\end{figure}

The setup introduced in this article includes multiple on-chip
beam splitters with varying distance. Examples are shown in
Fig.~\ref{beamsplitter2}. The conductor geometry defines the
separation, trajectories and finally the area of the
interferometer, enclosed by the waveguides. The bias field can
either be generated by external coils or by the conductors QP1 and
QP2. The setup in a) consists of two interferometers using either
the pair G1 + G5, or G2 + G4 and a bias field. In b) the
conductors G1 + G5 and  G2 + G4 form interferometers with the same
maximal separation of the atoms. However, due to the rapid
increase of the distance of the conductors the waveguides separate
faster. The conductor G3 realizes another interferometer. Here,
the current splits into two branches and is recombined at a
further position. Unbalance may occur if the current does not
split up equally.

\begin{figure}
  \centering
  \resizebox{\columnwidth}{!}{
  \includegraphics{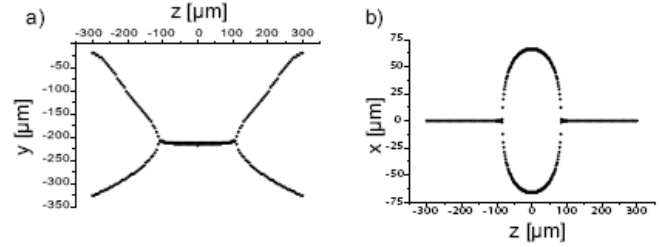}}
  \caption{Trajectories of the waveguide center for the interferometer shown in
  Fig.~\ref{beamsplitter2}{b}. The current in the conductors
  G2 and G4 is $I_{\mathrm{G2}}=I_{\mathrm{G4}}= 0.5$ A. An
  additional homogeneous bias field of $B_\perp = 5$ G is applied.
  a) and b) show projection onto the y-z and x-z planes, respectively.}
  \label{beamsplitter3}
\end{figure}
Fig.~\ref{beamsplitter3} shows trajectories of the waveguide
center for an interferometer formed by the conductors G2 and G4 in
Fig.~\ref{beamsplitter2} b). The current in each conductor is 0.5
A. A bias field of 5 G is applied. Fig. \ref{beamsplitter3} a)
shows a projection onto the y-z plane. Due to gravity, after
recombination a condensate will choose the lower branch for $ z <
-100\,\mu$m and $z> 100\,\mu$m.

Although the micro fabricated conductor pattern provides a well
defined geometry and precise control over magnetic fields, the
operation of these beam splitters is sensitive to ambient magnetic
stray fields. A magnetic field along the y-axis changes the
trajectories such that merging of the quadrupole waveguides is
inhibited. Instead, they pass each other. The adjustment of the
beam splitters may thus be nontrivial.

\section{Gravitational traps}

Let us consider the beam splitter with constant bias field and
varying distance between the conductors (section 3.2). The
variation of the distance between the conductors, as long as it is
smaller than the critical distance, only changes the height of the
waveguides above the chip. Provided that the chip is mounted
horizontally and the experiments are performed in gravity, a
gravitational potential arises. Potential wells and hills along
the waveguide can be achieved by changing the conductor geometry.

\subsection{Two wires with changing separation}

\begin{figure}[h]
  \centering
  \resizebox{\columnwidth}{!}{
  \includegraphics{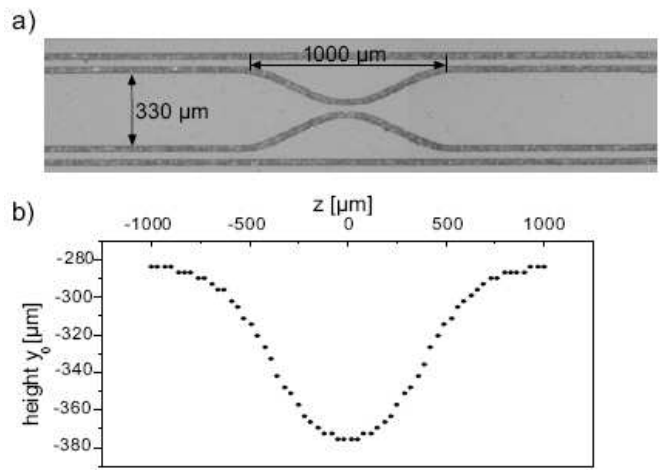}}
  \caption{Gravitational trap. a) Implementation of a gravitational trap for cold atoms on a chip.
  The magnetic field of the curved conductors in a bias field form a waveguide with
  varying height. b) Potential for a current of 1 A in the inner conductors and a
 homogeneous bias field of 10 G .}
  \label{gravtrap}
\end{figure}

A waveguide with a gravitational axial confinement can be realized
by two curved conductors and a bias field (Fig.~\ref{gravtrap}).
The waveguide forms with varying distance to the chip surface.
Calculating the potential for a current of 1 A in the curved
conductors and a bias field of 10 G (Fig.~\ref{gravtrap}~b) yields
that around its minimum $z = 0$, the potential can be approximated
by a parabola with a curvature of $b = 521\, \mathrm{Tm}^{-2}$ .
The height variation of $94\,\mu$m leads to a trap depth of $9.6
\,\mu$K, sufficient to trap a condensate or a thermal cloud near
the critical temperature. The overall distance of more than
300~$\mu$m to the chip surface assures that the potential of the
waveguide is not influenced by imperfections of the conductors. A
condensate could thus oscillate nearly unperturbed in such a trap.
The oscillation frequency can be calculated by considering the
potential energy:
\[
U = mgh = \frac{1}{2}mgbx^2 = \frac{1}{2}kx^2
\]
with $ k = mgb$. This leads to an harmonic oscillation with the
frequency
\[
\omega = \sqrt{\frac{k}{m}} = \sqrt{gb}.
\]

The oscillation frequency with the parameters given above is
$\omega = 2\pi \cdot 11.4$ Hz. These kind of traps could find use
in measurements of gravity since the oscillation frequency of a
condensate can be measured with high accuracy \cite{Ott2003}.

\subsection{One wire with changing width}

\begin{figure}[t]
  \centering
  \resizebox{\columnwidth}{!}{
  \includegraphics{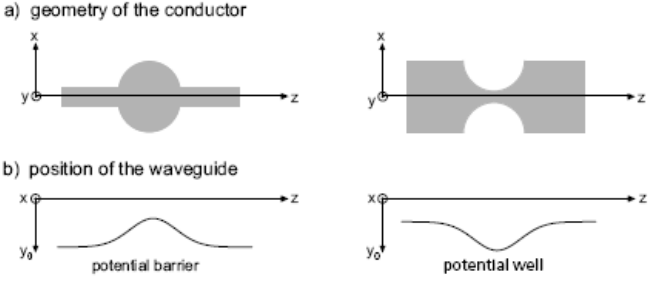}}
  \caption{Sketch of miniaturized gravitational traps. Broadening the conductor decreases the
  distance of the waveguide to the surface, narrowing increases it. In gravity, the axial
  potential of a waveguide exhibits gravitational barriers and wells. }
  \label{gravpotentials}
\end{figure}

Gravitational traps can be miniaturized by changing the width of the
conductor which is used for generating the waveguide. The geometry
of conductors for potential wells and barriers, as well as the
corresponding gravitational potentials, are plotted in
Fig.~\ref{gravpotentials}.
For the entire potential experienced by a cloud of paramagnetic
atoms, the magnetic field components due to changes in the conductor
geometry have to be taken into account.

\section{Conclusion}

We have demonstrated the fabrication of magnetic wave guides for
atoms with gold conductors on an aluminium oxide substrate. By
using a set of parallel conductors the confinement can be as high
as in standard optical traps even for large distances from the
surface where unwanted surface effects can be neglected. On the
same chip also spatial beam splitter geometries have been
realized. Finally, gravitational traps are demonstrated with an
axial oscillation frequency proportional to the square root of the
gravitational constant g.  Condensates in such traps offer novel
perspectives for the construction of gravimeters since they
oscillate practically frictionless with a very high Q-factor.

This work was supported by the Deutsche Forschungsgemeinschaft,
Landesstiftung Baden-W\"{u}rttemberg, and EU Marie-Curie RTN on
Atom Chips.

\end{document}